\documentclass[nofootinbib,aps,a4paper,superscriptaddress,showpacs]{revtex4}
\usepackage{amsmath}
\usepackage[dvips]{graphicx}
\begin{document}

\newcommand\be{\begin{equation}}
\newcommand\ee{\end{equation}}
\newcommand\bea{\begin{eqnarray}}
\newcommand\eea{\end{eqnarray}}
\newcommand\bseq{\begin{subequations}} %solo con amsmath
\newcommand\eseq{\end{subequations}}
\newcommand\bcas{\begin{cases}}
\newcommand\ecas{\end{cases}}
\newcommand{\p}{\partial}
\newcommand{\f}{\frac}

\title{Quantum Dynamics of the Taub Universe in a\\ Generalized Uncertainty Principle framework}

\author{Marco Valerio Battisti}
\email{battisti@icra.it}
\affiliation{ICRA - International Center for Relativistic Astrophysics}
\affiliation{Dipartimento di Fisica (G9), Universit\`a di Roma ``La Sapienza'' P.le A. Moro 5, 00185 Rome, Italy}
\author{Giovanni Montani}
\email{montani@icra.it} 
\affiliation{ICRA - International Center for Relativistic Astrophysics}
\affiliation{Dipartimento di Fisica (G9), Universit\`a di Roma ``La Sapienza'' P.le A. Moro 5, 00185 Rome, Italy}
\affiliation{ENEA C.R. Frascati (Dipartimento F.P.N.), Via Enrico Fermi 45, 00044 Frascati, Rome, Italy}
\affiliation{ICRANET C.C. Pescara, P.le della Repubblica 10, 65100 Pescara, Italy}

%\today

\begin{abstract}
The implications of a Generalized Uncertainty Principle on the Taub cosmological model are investigated. The model is studied in the ADM reduction of the dynamics and therefore a time variable is ruled out. Such a variable is quantized in a canonical way and the only physical degree of freedom of the system (related to the Universe anisotropy) is quantized by means of a modified Heisenberg algebra. The analysis is performed at both classical and quantum level. In particular, at quantum level, the motion of wave packets is investigated. The two main results obtained are as follows. i) The classical singularity is probabilistically suppressed. The Universe exhibits a stationary behavior and the probability amplitude is peaked in a determinate region. ii) The GUP wave packets provide the right behavior in the establishment of a quasi-isotropic configuration for the Universe.  
\end{abstract}


\maketitle 

\section{Introduction}

The existence of a fundamental minimal scale has long been expected in a quantum theory of gravity. In fact, from basic considerations, the combination of (general) relativistic and quantum effects implies that the conventional picture of a smooth spacetime breaks down at some fundamental scale. An immediate way to realize this intuition can be found in the so-called Generalized Uncertainty Principle (GUP). In fact, if one considers some corrections to the usual Heisenberg uncertainty principle, a non-zero minimal-length uncertainty naturally appears.

Interest in minimal length or Generalized Uncertainty Principle has been motivated by studies in perturbative string theory \cite{String}, considerations on the proprieties of black holes \cite{Mag} and de Sitter space \cite{Sny}. In particular, from the string theory point of view, a minimal observable length it is a consequence of the fact that strings can not probe distance below the string scale. However, in recent years, a big amount of work has been done in this active field in a wide variety of directions (see for example \cite{GUP1} and the references therein; for another application of the GUP approach to the minisuperspace dynamics, from a different point of view, see \cite{Vakili}). In particular, in our previous work \cite{BM07}, the flat isotropic cosmological model was analyzed in this framework. It was shown how the Universe dynamics is singularity free, but no evidence about a Big-Bounce arises. In fact, in such a case, the quantum Universe approaches a stationary state ``near the Planckian region''.

In this work we will consider the Taub Universe in the GUP framework. This cosmological model is of particular interest because it is a special case of the Bianchi IX model, i.e. the most general (together with Bianchi VIII) homogeneous cosmological model. The approach to the singularity of a Bianchi IX model is described by a particle in two dimension (the two physical degree of freedom of the Universe, i.e. the anisotropy variables) moving in a potential with exponential walls bounding a triangle. Such a particle is reflected by the walls and the dynamics appears to be chaotic. Taub model is the particular case of Bianchi IX, only consisting of a one potential wall (one of the anisotropy variables is taken to be zero). Therefore, classically, the particle bounces against this wall and then is reflected toward the classical singularity. In the billiard representation, as we will see, the Taub model is a part of this cosmological billiard and no chaotic dynamic appears. Therefore, the study of the Taub Universe is a necessary step toward the analysis of the Mixmaster Universe in the GUP approach.

Let us discuss some aspects of the application of the GUP framework in quantum cosmology. By the minisuperspace reduction, a genuine quantum field theory (quantum general relativity) reduces to a quantum mechanical system (homogeneous quantum cosmology). As well-known the homogeneous Universes, in the vacuum case, are characterized by only three degrees of freedom. Thus, an homogeneous model is nothing but a three-dimensional mechanical system. In this respect, the GUP approach to quantum cosmology appears physically grounded. In fact, a generalized uncertainty principle can be immediately reproduced modifying the canonical Heisenberg algebra. Although such a deformed commutation relation, differently from the GUP itself, has not been so far derived directly from string theory, it is one possible way in which certain features of a more fundamental theory may manifest themselves in some toy models (finite degrees of freedom). This way, the GUP approach relies on a modification of the canonical quantization prescriptions and, in this respect, it can be reliably applied to any dynamical system. Thus, in the application to minisuperspace models, such a scheme appears the most natural with respect the other ``non-commutative approaches''. Therefore, the GUP formalism allows us to implement some peculiar features of a more general theory, for example string theory, in quantizing a cosmological model. 

The Taub model will be studied in the context of the ADM reduction of the dynamics. Such a representation, allows us to regard one variable, mainly the Universe volume, as a ``time'' for the dynamics. Therefore, such a Universe will be described by the motion of a one-dimensional particle in a domain limited by a potential wall. This variable, describing the physical degree of freedom of the system, will be treated in the GUP formalism, while the time-variable in a canonical way. The analysis in the GUP formalism will be performed at both classical and quantum level, obtaining a radically different dynamical picture from the canonical case. At classical level, the trajectories of the particle before and after the bounce at the potential wall will be much closer to each other than in the canonical case. However, the most interesting results appear at quantum level. In fact, the motion of the wave packets in both, Wheeler-DeWitt (WDW) and GUP approaches is analyzed. In the canonical case (WDW theory), the wave packets are peaked around the classical trajectories and, after the bounce on the potential wall, they fall in the cosmological singularity. On the other hand, in the GUP case, we obtain two remarkable results. i) The probability density to find the Universe is peaked ``near'' the potential wall and the wave packets show a stationary behavior. Therefore, the classical singularity will be not probabilistically privileged. ii) The value of anisotropy for which the probability amplitude is peaked corresponds to a quasi-isotropic Universe. Therefore, the GUP wave packets exhibit the right behavior in predicting an isotropic Universe.

The paper is organized as follows. Sections II and III are devoted to a brief review on the fundamental aspects of the Taub Universe and the formulation of a quantum mechanics with a non-zero minimal uncertainty in position, respectively. In Section IV, the ``deformed'' classical analysis of the dynamics is shown. Then, in Section V, the quantization of the model, according to the GUP approach, is performed and the comparison between the WDW and the GUP wave packets is explained in detail. Finally in Section VI, the semiclassical limit of the model is discussed. Concluding remarks follow.

Over the whole paper we adopt units such that $\hbar=c=16\pi G=1$.

\section{The model}

The Bianchi cosmological models \cite{RS} are spatially homogeneous models such that the symmetry group acts {\it simply transitively}\footnote{Let $G$ a Lie group, $G$ is said to act {\it simply transitively} on the spatial manifold $\Sigma$ if, for all $p,q\in\Sigma$, there is a unique element $g\in G$ such that $g(p)=q$.} on each spatial manifold. The Bianchi IX model, together with Bianchi VIII, is the most general one and its line element reads, in the Misner parametrization \cite{Mis69}, 
\be
ds^2=N^2dt^2-e^{2\alpha}\left(e^{2\gamma}\right)_{ij}\omega^i\otimes\omega^j,
\ee
where $N=N(t)$ is the lapse function and the left invariant 1-forms $\omega^i=\omega^i_adx^a$ satisfy the Maurer-Cartan equation $2d\omega^i=\epsilon^i_{jk}\omega^j\wedge\omega^k$. The variable $\alpha=\alpha(t)$ describes the isotropic expansion of the Universe and $\gamma_{ij}=\gamma_{ij}(t)$ is a traceless symmetric matrix $\gamma_{ij}=diag\left(\gamma_++\sqrt3\gamma_-,\gamma_+-\sqrt3\gamma_-,-2\gamma_+\right)$ which determines the shape change (the anisotropy) {\it via} $\gamma_\pm$. Since the determinant of the 3-metric is given by $h=\det e^{\alpha+\gamma_{ij}}=e^{3\alpha}$, the classical singularity appears for $\alpha\rightarrow-\infty$. Performing the usual Legendre transformations, we obtain the Hamiltonian constraint for this model. As well-known \cite{Mis69,CGM} the dynamics of the Universe, toward the singularity, is described by the motion of a two-dimensional particle (the two physical degree of freedom of the gravitational field) in a dynamically-closed domain. In the Misner picture, such a domain depends on the time variable $\alpha$. To overcame this difficulty, the so-called Misner-Chitr$\acute e$-like variables \cite{Chi} are introduced {\it via}
\be\label{txt}
\alpha=-e^\tau\xi, \qquad \gamma_+=e^\tau\sqrt{\xi^2-1}\cos\theta, \qquad \gamma_-=e^\tau\sqrt{\xi^2-1}\sin\theta
\ee
with $\xi\in[1,\infty)$ and $\theta\in[0,2\pi]$. The goal for the use of these variables relies on the fact that the dynamically-allowed domain becomes independent of $\tau$, which will behave as the time variable. In terms of these new variables, the Hamiltonian constraint rewrites  
\be
H=-p_\tau^2+p_\xi^2(\xi^2-1)+\f{p_\theta^2}{\xi^2-1}\approx0.
\ee
\begin{figure}
\begin{center}
\includegraphics[height=2in]{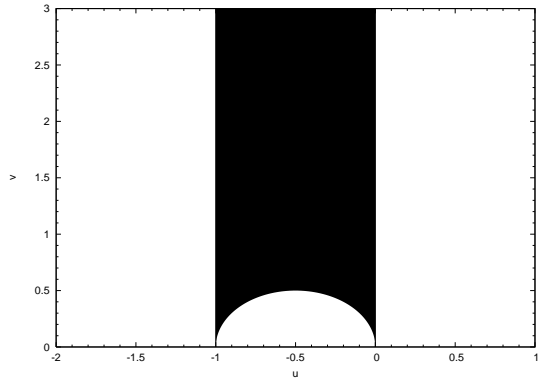}
\caption{The dynamical-allowed domain $\Gamma_Q(u,v)$ in the Poincar$\acute e$ complex upper half-plane where the dynamics of the Universe is restricted, toward the classical singularity, by the potential.} 
\end{center}
\end{figure}
Let us perform the ADM reduction of the dynamics. This scheme relies on the idea to solve the classical constraint, with respect to a given momenta, before implementing some quantization algorithm. In this way, we will obtain an effective Hamiltonian which will depend only on the physical degrees of freedom of the system. Therefore we solve the constraint $H=0$ with respect to $p_\tau$, and thus, from now on, we will consider the variable $\tau$ as the time coordinate for the dynamics (we take the time gauge $\dot\tau=1$), obtaining
\be\label{hxt}
-p_\tau=\sqrt{p_\xi^2(\xi^2-1)+\f{p_\theta^2}{\xi^2-1}}.
\ee
The dynamics of such a system is equivalent to a billiard ball on a Lobatchevsky plane \cite{ChBa83}, as we can see making use of the Jacobi metric\footnote{This approach reduces the equations of motion of a generic system to a geodesic problem on a given manifold.}. With another coordinate transformation $(\xi,\theta)\rightarrow(u,v)$, it is possible to choose the so-called Poincar$\acute e$ representation in the complex upper half-plane \cite{KM97}. Such new variables are defined as
\be\label{uv}
\xi=\f{1+u+u^2+v^2}{\sqrt3v}, \qquad \theta=-\tan^{-1}\left(\f{\sqrt3(1+2u)}{-1+2u+2u^2+2v^2}\right),
\ee
and the dynamical-allowed domain $\Gamma_Q=\Gamma_Q(u,v)$ is plotted in Fig. 1. We note how the three corners in the Misner picture are replaced by the points $(0,0)$, $(-1,0)$ and $v\rightarrow\infty$ in the $(u,v)$-plane. The ADM ``constraint'', in this $(u,v)$ scheme, is simpler than the previous one (\ref{hxt}) and becomes
\be\label{huv}
-p_\tau\equiv H_{ADM}=v\sqrt{p_u^2+p_v^2}.
\ee 
In order to simplify the model we will consider the case $\gamma_-=0$, which corresponds to the so-called Taub model \cite{RS}. The dynamics of this Universe is equivalent to the motion of a particle in a one-dimensional closed domain. Such a domain corresponds to taking only one of the three equivalent potential walls of the Bianchi IX model. As we can see from (\ref{txt}) and (\ref{uv}), this particular case appears for $\theta=0\Rightarrow u=-1/2$ ($\xi=(v^2+3/4)/\sqrt3v$) and the ADM Hamiltonian (\ref{huv}) rewrites 
\be\label{hv}
H_{ADM}^T=vp_v,
\ee
being $v\in[1/2,\infty)$, as shown in Fig. 1. The above Hamiltonian (\ref{hv}) can be further simplified defining a new variable $x=\ln v$ and becomes
\be\label{ht}
H_{ADM}^T=p_x\equiv p,
\ee
which will be the starting point of our analysis. Let us stress that the classical singularity now appears for $\tau\rightarrow\infty$.

\section{Quantum mechanics in the GUP framework}

In this Section we briefly review some aspects and results of a non-relativistic quantum mechanics with non-zero minimal uncertainties in position \cite{Kem}. In one dimension, we consider the Heisenberg algebra generated by $\bf q$ and $\bf p$ obeying the commutation relation  
\be\label{modal}
[{\bf q},{\bf p}]=i(1+\beta{\bf {p}}^2), 
\ee
where $\beta$ is a ``deformation'' parameter. This commutation relation leads to the uncertainty relation
\be\label{gup}
\Delta q \Delta p\geq \f 1 2\left(1+\beta (\Delta p)^2+\beta \langle{\bf p}\rangle^2\right),
\ee
which appears in perturbative string theory \cite{String}. The canonical Heisenberg algebra can be recovered in the limit $\beta=0$ and the generalization to more dimension is straightforward, leading naturally to a ``noncommutative geometry'' for the space coordinates.

It is immediate to verify that such a Generalized Uncertainty Principle (\ref{gup}) implies a finite minimal uncertainty in position $\Delta q_{min}=\sqrt\beta$. As well-known, the existence of a non-zero uncertainty in position implies that there cannot be any physical state which is a position eigenstate. In fact, an eigenstate of an observable necessarily has vanishing uncertainty on it. To be more precise, let us assume the commutation relations to be represented on some dense domain $D\subset\mathcal H$ in a Hilbert space $\mathcal H$. In the canonical case, a sequence $\vert\psi_n\rangle\in D$ with position uncertainties decreasing to zero exists. On the other hand, in presence of a minimal uncertainty $\Delta q_{min}\geq0$, it is not possible any more to find some $\vert\psi_n\rangle\in D$ such that
\be
\lim_{n\rightarrow\infty}\left(\Delta q_{min}\right)_{\vert\psi_n\rangle}=\lim_{n\rightarrow\infty}\langle\psi\vert({\bf q}-\langle\psi\vert{\bf q}\vert\psi\rangle)^2\vert\psi\rangle=0.
\ee 
Although it is possible to construct position eigenvectors, they are only formal eigenvectors and not physical states. Let us now stress that this feature comes out from the corrections to the canonical commutation relations and, in general, a non-commutativity of the ${\bf q}_i$ will not imply a finite minimal uncertainty $\Delta q_{min}\geq0$. Therefore, in the GUP approach, we cannot work in the configuration space and the notion of {\it quasiposition} will recovered in what follows. 

The Heisenberg algebra (\ref{modal}) can be represented in the momentum space, where the $\bf q$, $\bf p$ operators act as 
\be\label{rep}
{\bf p}\psi(p)=p\psi(p), \qquad {\bf q}\psi(p)=i(1+\beta p^2)\p_p\psi(p),
\ee
on a dense domain $S$ of smooth functions. The measure in the scalar product on this domain, with respect to which $\bf q$ and $\bf p$ are symmetric, results to be $dp/(1+\beta p^2)$. To recover information on positions we have to study the states that realize the maximally-allowed localization. Such states $\vert\psi^{ml}_{\zeta}\rangle$ of maximal localization, which are proper physical states around a position $\zeta$, have the proprieties $\langle\psi^{ml}_{\zeta}\vert {\bf q}\vert\psi^{ml}_{\zeta}\rangle=\zeta$ and $(\Delta q)_{\vert\psi^{ml}_{\zeta}\rangle}=\Delta q_{min}$. These states are called of maximal localization, because they obey the minimal uncertainty condition $\Delta q\Delta p=\vert\langle [{\bf q},{\bf p}]\rangle\vert/2$ and therefore the following equation holds
\be
\left({\bf q}-\langle{\bf q}\rangle + \f{\langle [{\bf q},{\bf p}]\rangle}{2(\Delta p)^2}({\bf p}-\langle{\bf p}\rangle)\right){\vert\psi^{ml}_{\zeta}\rangle}=0,
\ee
which admit, in the momentum space, the following solution\footnote{The absolutely minimal uncertainty in position $\Delta q_{min}=\sqrt\beta$ and thus also the maximal localization states, are obtained for $\langle{\bf p}\rangle=0$.}
\be
\psi^{ml}_{\zeta}(p)\sim\f 1 {(1+ \beta p^2)^{1/2}} \exp\left(-i\f{\zeta}{\sqrt{\beta}} \tan^{-1}(\sqrt{\beta}p)\right),
\ee
where with $\sim$ we omit the normalization constant. As we can easily see these states, in the $\beta=0$ limit, reduce to ordinary plane waves. An important difference with the canonical case relies on the fact that such maximal-localization states are normalizable and therefore their scalar product is a function rather then a distribution (the Dirac $\delta$-distribution). As last step, we can project an arbitrary state $\vert\psi\rangle$ on the maximally localized states $\vert\psi^{ml}_{\zeta}\rangle$ in order to obtain the probability amplitude for a particle being maximally localized around the position $\zeta$ (i.e. with standard deviation $\Delta q_{min}$). We call these projections the ``quasiposition wave function'' $\psi(\zeta)\equiv\langle\psi^{ml}_{\zeta}\vert\psi\rangle$; explicitly, we have
\be\label{qwf} 
\psi(\zeta)\sim\int^{+\infty}_{-\infty}\f{dp}{(1+\beta p^2)^{3/2}} \exp\left(i\f{\zeta}{\sqrt{\beta}} \tan^{-1}(\sqrt{\beta}p)\right)\psi(p).
\ee
This is nothing but a generalized Fourier transformation, where in the $\beta=0$ limit the ordinary position wave function $\psi(\zeta) = \langle\zeta\vert\psi\rangle$ is recovered.

\section{Generalized classical dynamics}

Before quantizing the model (described in the previous Section), in the GUP framework, let us now perform a ``deformed'' classical analysis. To be more precise, we will check the modification arising from the deformed Heisenberg algebra (\ref{modal}) on the classical trajectory of the Universe described by the Hamiltonian (\ref{ht}). Such ``deformed'' classical dynamics is contained in the modified symplectic geometry arising from the classical limit of (\ref{modal}), as soon as the parameter $\beta$ is regarded as an independent constant with respect $\hbar$. Therefore, it is possible to replace the quantum-mechanical commutator (\ref{modal}) {\it via} the Poisson bracket
\be\label{pm}
-i[{\bf q},{\bf p}]\Longrightarrow\{{\bf q},{\bf p}\}=(1+\beta p^2).
\ee 
From a string theory point of view, keeping the parameter $\beta$ fixed as $\hbar\rightarrow0$ corresponds to keeping the string momentum scale fixed while the string length scale shrinks to zero. Also in this case, the generalization to more dimensions is straightforward and was performed in \cite{Ben} to study the effects of the ``deformation'' on the classical orbits of particles in a central force potential.

Therefore, from (\ref{pm}), the Poisson bracket for any two-dimensional phase space function appears to be
\be
\{F,G\}=\left(\f{\p F}{\p q}\f{\p G}{\p p}-\f{\p F}{\p p}\f{\p G}{\p q}\right)(1+\beta p^2),
\ee
and thus the equations of motion read
\be\label{eqmod}
\dot q=\{q,H\}=(1+\beta p^2)\f{\p H}{\p p}, \qquad \dot p=\{p,H\}=-(1+\beta p^2)\f{\p H}{\p q}.
\ee 
Applying equations (\ref{eqmod}) to our Hamiltonian (\ref{ht}), in which case the $\dot F$ denotes the ``time'' derivative $dF/d\tau$, we immediately obtain
\be\label{moteq}
x(\tau)=(1+\beta A^2)\tau+cost, \qquad p(\tau)=cost=A,
\ee
where we remember that $x\in[\ln(1/2),\infty)$. Therefore the effects, at a classical level, of the deformed Heisenberg algebra (\ref{modal}), i.e. the GUP approach, on the Taub Universe are as follows. As we can easily see from the equations of motion (\ref{moteq}), the angular coefficient is $(1+\beta A^2)>1$ for $\beta\neq0$. In other words, the angle between the two straight lines $x(\tau)$, for $\tau<0$ and $\tau>0$, becomes smaller as the values of $\beta$ grows. Thus the trajectories of the particle (Universe), before and after the bounce on the potential wall at $x=x_0\equiv\ln(1/2)$, are ``closer'' to each other then in the canonical case ($\beta=0$). In fact, the incoming trajectory (for $\tau<0$) will be closer, for $\beta\neq0$, to the outgoing trajectory (for $\tau>0$) toward the classical singularity.

\section{Generalized quantization of the model}

In this Section we focus our attention on the generalized quantum features of the Taub Universe, described by the Hamiltonian (\ref{ht}) with the condition $x\in[x_0,\infty)$. Namely, we perform a generalized quantization of this model based on the GUP formalism reviewed in the previous paragraph. Let us stress that, from the ADM reduction of the dynamics, the variable $\tau$ is regarded as a time coordinate and therefore the conjugate couple ($\tau,p_\tau$) will be treated in a canonical way. This way, we deal with a Schr\"odinger-like equation
\be
i\p_\tau\Psi(\tau,p)=\hat H_{ADM}^T\Psi(\tau,p).
\ee
Therefore, applying the algebra representation (\ref{rep}) to the model, we obtain the eigenvalue problem 
\be\label{eqaut}
k^2\psi_k(p)=p^2\psi_k(p),
\ee
where 
\be
\Psi_k(\tau,p)=\psi_k(p)e^{-ik\tau}.
\ee
We have to square the eigenvalue problem in order to correctly impose the boundary condition. Such an operation poses no problem. In fact, in agreement with the analysis developed in \cite{Puzio}, we make the well-grounded hypothesis that the eigenfunctions form is independent of the presence of the square root, since its removal implies the square of the eigenvalues only. 

As we have explained, in the GUP approach, we have lost all informations on the position itself. Therefore, the boundary condition have to be imposed on the ``quasiposition wave function'' (\ref{qwf}), in the sense that $\psi(\zeta_0)=0$ (being $\zeta_0=\langle\psi^{ml}_{\zeta}\vert x_0\vert\psi^{ml}_{\zeta}\rangle$ in agreement with the previous discussion). The solution of the above equation (\ref{eqaut}) is the Dirac $\delta$-distribution 
\be
\psi_k(p)=\delta(p^2-k^2)
\ee
and therefore the ``quasiposition wave function'' (\ref{qwf}) reads 
\begin{multline} 
\psi_k(\zeta)\sim\int^{+\infty}_{-\infty}\f{dp}{(1+\beta p^2)^{3/2}} \exp\left(i\f{\zeta}{\sqrt{\beta}} \tan^{-1}(\sqrt{\beta}p)\right)\f1k\left(A \delta(p-k)+B\delta(p+k)\right)=\\=\f1{k(1+\beta k^2)^{3/2}}\left[A\exp\left(i\f{\zeta}{\sqrt{\beta}} \tan^{-1}(\sqrt{\beta}k)\right)+B\exp\left(-i\f{\zeta}{\sqrt{\beta}} \tan^{-1}(\sqrt{\beta}k)\right)\right],
\end{multline}
where $A$ and $B$ are integration constants. In this way, the boundary condition $\psi(\zeta_0)=0$ can be easily imposed and fixes one constant, giving us the final form for the ``quasiposition'' eigenfunctions
\be\label{ef}
\psi_k(\zeta)=\f A{k(1+\beta k^2)^{3/2}}\left[\exp\left(i\f{\zeta}{\sqrt{\beta}} \tan^{-1}(\sqrt{\beta}k)\right)-\exp\left(i\f{(2\zeta_0-\zeta)}{\sqrt{\beta}} \tan^{-1}(\sqrt{\beta}k)\right)\right].
\ee

To get a better feeling of our quantum Universe we construct and examine the motion of wave packets. In fact, the analysis of dynamics of such wave packets allow us to give a precise description of the evolution of the Taub model. Such an evolution will be preformed in either GUP and WDW approaches (the last one regarded as a limiting case of the GUP scheme) and reliable differences between these two theories will appear. The wave packets are superposition of the eigenfunctions, i.e.
\be\label{wapa}
\Psi(\tau,\zeta)=\int_0^\infty dk A(k)\psi_k(\zeta)e^{-ik\tau}.
\ee
In the following we will take $A(k)$ as a Gaussian-like function
\be\label{gau}
A(k)=k(1+\beta k^2)^{3/2}e^{-\f{(k-k_0)^2}{2\sigma^2}}
\ee
in order to simplify the explicit expression of the wave packets. The computation of (\ref{wapa}) for the eigenfunctions (\ref{ef}) is performed in a numerical way and the parameters are chosen as follows: $k_0=1$ and $\sigma=4$.
\begin{figure}
\includegraphics[height=2in]{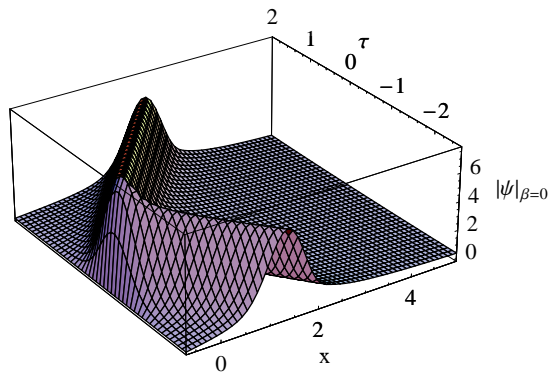} 
\caption{The evolution of the wave packets $\vert\Psi(\tau,x)\vert$ in the canonical case, i.e. $\beta=0$. The $x$ variable is in the $[x_0,5]$-interval.} 
\end{figure}

\subsection{Predictions of the WDW approach}

Let us analyze the GUP modifications induced to the canonical evolution of such wave packets. As we have said, in the $\beta=0$ limit, we recover the canonical scheme. Therefore, first of all, we discuss the motion of wave packets (\ref{wapa}) in the canonical approach, i.e. in the Wheeler-DeWitt formalism. In such a limit, the eigenfunctions (\ref{ef}) reduce to ordinary plane waves and the quasiposition $\zeta\rightarrow x$. The superposition of these eigenfunctions, with the Gaussian-like weight functions (\ref{gau}), can be computed analytically and the result is plotted in Fig. 2. As we can see from the picture, the wave packets follow the classical trajectories described in the previous Section in the $\beta=0$ case. The probability amplitude to find the particle (Universe) is packed around these trajectories. In this respect no privileged regions arise, namely no dominant probability peaks appear in the ($\tau,x$)-plane. As matter of fact, the ``incoming'' Universe ($\tau<0$) bounce at the potential wall at $x=x_0$ and then fall toward the classical singularity ($\tau\rightarrow\infty$). Therefore, as well-known, the Wheeler-DeWitt formalism is not able to get light on the necessary quantum resolution of the classical cosmological singularity. As we will see in a while, this picture is radically changed in the GUP framework.   

\subsection{Predictions of the GUP approach}

As we said the parameter $\beta$, i.e. the presence of a non-zero minimal uncertainty in the configuration variable, is responsible for the GUP effects on the dynamics. In our case the variable $x=\ln v$, which is related via the relation (\ref{txt}) for $\theta=0$ and $\xi=(v^2+3/4)/\sqrt3v$ to the Universe anisotropy $\gamma_+$ as
\be\label{anix}
\gamma_+=\f{e^\tau}{\sqrt3v}\left(v^2-\f34\right)=\f{e^{\tau-x}}{\sqrt3}\left(e^{2x}-\f34\right),
\ee
is the configuration variable for the system. Therefore, since the relation $\Delta x_{min}=\sqrt\beta$ appears, we see that the physical interpretation of $\beta$ is to give a non-zero minimal uncertainty in the anisotropy of the Universe. Thus, in order to comprehend the alterations induced by the deformed Heisenberg algebra on the canonical Universe dynamics, we have to analyze different $\beta$-regions. In fact, when the ``deformation'' parameter $\beta$ becomes more and more important, i.e. when we are at some scale which allows us to appreciate the GUP effects, the evolution of the wave packets is different from the canonical case. To be more precise, to see these results what really matters is when the product $k_0\sqrt\beta$ becomes remarkable, i.e. when $k_0\sqrt\beta\sim\mathcal O(1)$ and therefore when $\beta$ is comparable to $1/k_0^2$. In fact, as explained in the next Section, it is only necessary the wave packets (\ref{wapa}) peak at energies much smaller then $1/\sqrt\beta$ in order to recover the correct semiclassical behavior of the model far from the singularity. In particular, for $k_0=1$ we can distinguish between three different $\beta$-regimes (see Fig. 3):
\begin{itemize}
	\item Let us first consider the ($\beta\sim\mathcal O(10^{-2})$)-region. In this regime the wave packets begin to spread and a constructive and destructive interference between the incoming and outgoing wave appears. The probability amplitude to find the Universe is still peaked around the classical trajectory, but ``not so much'' as in the canonical case.
\end{itemize}
 
\begin{itemize}
	\item When this parameter becomes more influent, i.e. $\beta\sim\mathcal O(10^{-1})$, we can no more distinguish an incoming or outgoing wave packet. At this level, speaking about a wave packet which follows the classical trajectory is meaningless. Moreover, the probability amplitude to find the Universe is, in some sense, peaked in a specified region in the ($\tau,\zeta$)-plane, i.e. for $\zeta\simeq0$.
\end{itemize}
 
\begin{itemize}
	\item As last step, for $\beta\sim\mathcal O(1)$, a dominant probability peak ``near'' the potential wall appears. In this $\beta$-region, there are also other small peaks for growing values of $\zeta$, but they are widely suppressed for bigger $\beta$. In this case, the motion of wave packets shows a stationary behavior, i.e. these are independent of $\tau$.
\end{itemize}
Following this picture we are able to learn the GUP modifications to the WDW wave packets evolution. In fact, considering a sort of dynamics in the ``deformation'' parameter $\beta$, from small to ``big'' values of $\beta$, we can see how the wave packets ``escape'' from the classical trajectories and approach a stationary state close to the potential wall. All this picture is plotted in Fig. 3. 
\begin{figure}
\includegraphics[height=2in]{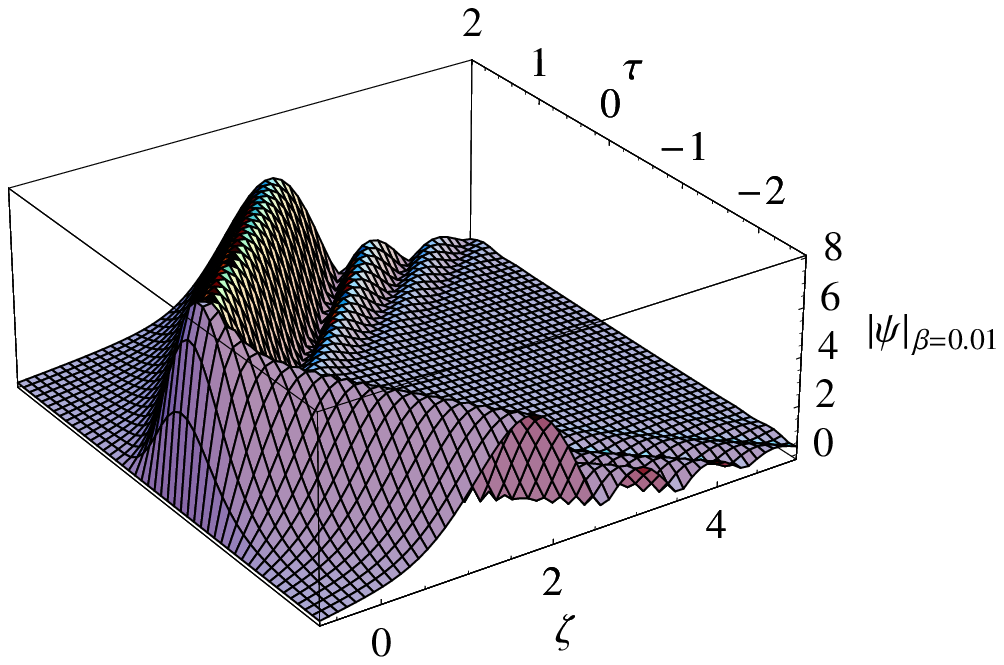} \qquad \qquad \qquad \includegraphics[height=2in]{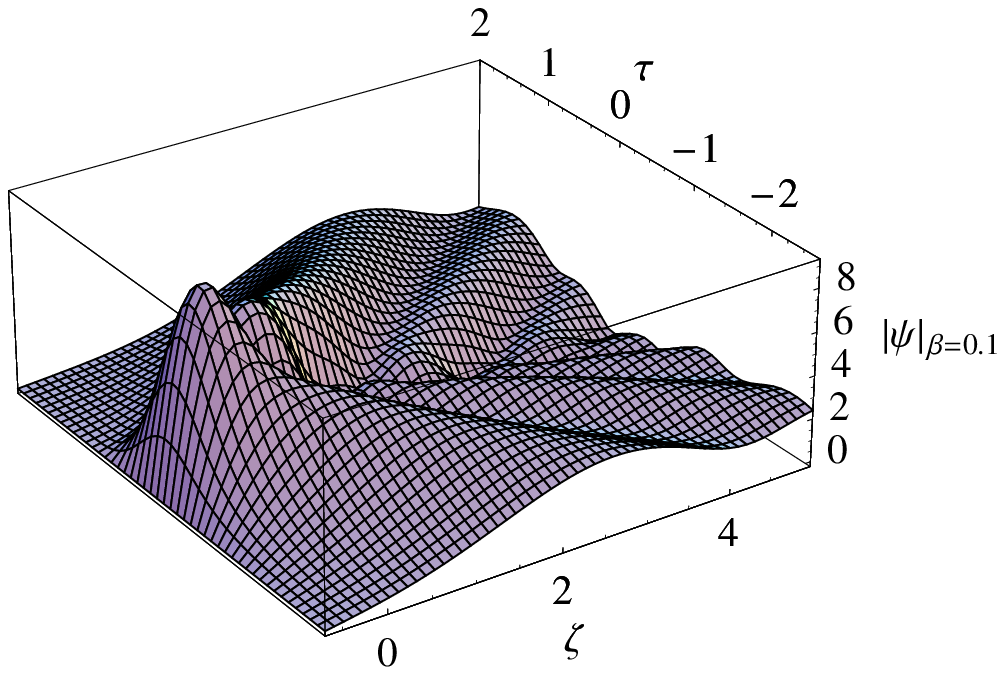} \\ \includegraphics[height=2in]{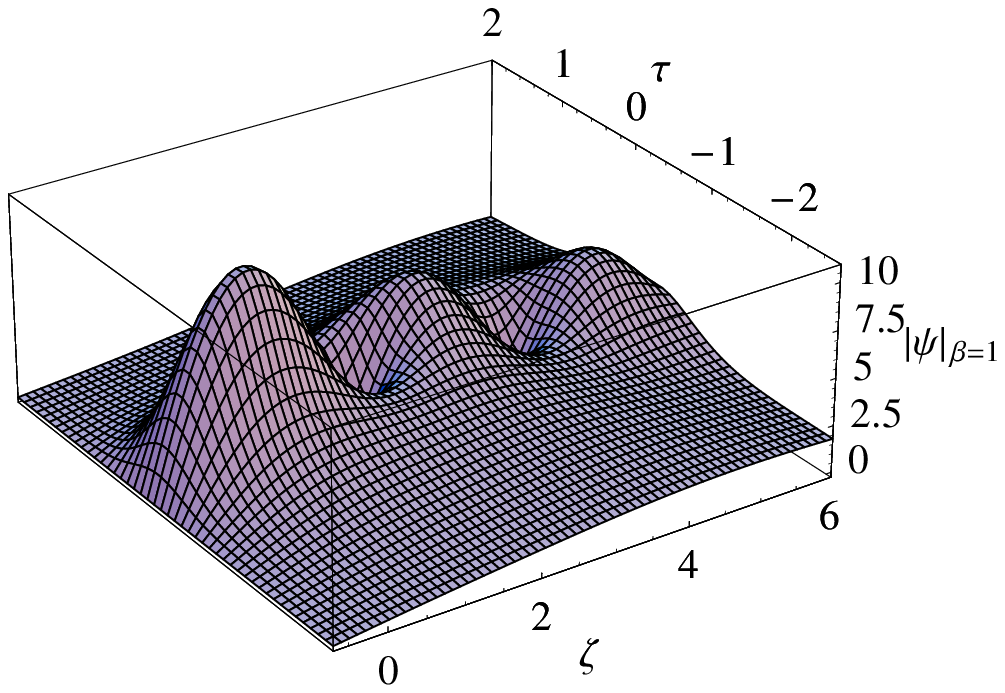}
\caption{The evolution of the wave packets $\vert\Psi(\tau,\zeta)\vert$ in the GUP framework. The graphics are for $\beta k_0^2=0.01$, $\beta k_0^2=0.1$ and $\beta k_0^2=1$ respectively. For smaller $\beta k_0^2$ the canonical case is recovered. In particular in this plot we choose $k_0=1$.} 
\end{figure}

Such a behavior, in some sense, is expected from a classical point of view. In fact, we have seen how the consequences of the GUP approach, at the classical level, reside in the shrinking of the ingoing and the outgoing trajectories on the potential wall. So a quantum probability interference, which is present in the ($\beta\sim\mathcal O(10^{-2})$)-region is a fortiori predicted. On the other hand, the stationarity feature exhibited by the Universe in the ($\beta\sim\mathcal O(1)$)-region, is a purely quantum GUP effect. In fact, such a behavior cannot be inferred from a ``deformed'' classical analysis. From this point of view, the classical singularity ($\tau\rightarrow\infty$) is widely probabilistically suppressed, because the probability to find the Universe is peaked just around the potential wall. On the other hand, in the WDW approach, as we said, the wave packets follow the classical trajectory toward the singularity. 

Another important feature to be considered is that the large anisotropy states are not privileged. In fact the most probable ones, as we can see from Fig. 3, are those for $\tau\simeq\zeta\simeq0$, i.e. from equation (\ref{anix}) those for which\footnote{We remember that, according with the previous discussion, $\zeta=\langle\psi^{ml}_{\zeta}\vert {\bf x}\vert\psi^{ml}_{\zeta}\rangle$. Therefore, we assume that the relation (\ref{anix}) is still valid at the quantum level replacing $x$ with the ``quasiposition'' $\zeta$.} $|\gamma_+|\simeq\mathcal O(10^{-1})$. Therefore, the GUP wave packets favor the establishment of a quantum quasi-isotropic configuration for the Universe differently from those in the WDW theory.

\section{Physical considerations and the semiclassical limit}

To complete the study of the quantum dynamics of the Taub Universe in the GUP framework, let us now discuss some physical aspects of the model illustrating how it exhibits the correct semiclassical limit away from the {\it cut-off} region. First, we have to analyze the geometrical (physical) meaning of the adopted variables. By equation (\ref{anix}), a monotonic relation between the anisotropy of the Universe $\gamma_+$ and our (classical) configuration variable $x=\ln v\in[x_0\equiv\ln(1/2),\infty)$ appears and therefore, the variable $x$ can be regarded as a measure of the model anisotropy. In particular, the isotropic shape of the Taub Universe ($\gamma_+=0$) comes out for a particular value of $x$, i.e. $x=\ln(\sqrt3/2)$ and we get the closed Friedmann-Robertson-Walker Universe. 

We stress that at the GUP quantum level, as explained in Section III, we lost direct information on the position itself and the knowledge on it can be recovered only by the study of the states which realize the maximally-allowed localization. In particular, the ordinary (WDW) position wave function $\psi(x) = \langle x\vert\psi\rangle$ is replaced by the ``quasiposition wave function'' (\ref{ef}) $\psi(\zeta)=\langle\psi^{ml}_{\zeta}\vert\psi\rangle$, where $\zeta$ is defined by $\langle\psi^{ml}_{\zeta}\vert {\bf x}\vert\psi^{ml}_{\zeta}\rangle=\zeta$ and $(\Delta x)_{\vert\psi^{ml}_{\zeta}\rangle}=\Delta x_{min}=\sqrt\beta$. Of course in the $\beta=0$ limit, or more correctly for momenta conjugate to $x$ much less than the {\it cut-off} value (i.e. for $p\ll1/\sqrt\beta$, $k\ll1/\sqrt\beta$), the canonical framework is recovered, i.e. the GUP wave function $\psi(\zeta)$ gives the WDW wave function $\psi(x)$ 
\be
\psi^{GUP}(\zeta)=\langle\psi^{ml}_{\zeta}\vert\psi\rangle\rightarrow\psi^{WDW}(x) = \langle x\vert\psi\rangle, \qquad \zeta\rightarrow x.
\ee
Let us stress once more that the ``quasiposition wave function'' $\psi(\zeta)$ represents the probability amplitude for the particle (Universe) being maximally localized around the position $\zeta$, i.e. with the standard deviation $\sqrt\beta$.

In this work, the GUP scheme was applied to the configuration variable $x$ which describes the only physical degree of freedom of the Taub model {\it via} the relation (\ref{anix}). On the other hand, the time variable $\tau$ and its conjugate momenta $p_\tau$ (namely $k$), was treated in the canonical picture. This way, we have imposed a {\it cut-off} in the model by the non-vanishing minimal uncertainty on the position $x$, coming out from the deformed Heisenberg algebra (\ref{modal}). In this respect, the GUP effects on the Universe dynamics are not directly related to the time variable which describes the evolution toward the singularity, appearing for $\tau\rightarrow\infty$. 

According to the previous discussion, the semiclassical Taub dynamics, i.e. the WDW evolution described in Section VA (see Fig. 2), is obtained analyzing the dynamics of a wave packet peaked at energies much smaller with respect to the {\it cut-off} ones. More precisely, we have to take the weight function (\ref{gau}) peaked at $k_0\ll1/\sqrt\beta$ and to evaluate the evolution of the wave packet (\ref{wapa}) far from the singularity, i.e. for $\tau\rightarrow-\infty$. As a matter of fact, considering the Universe in the ($\beta\sim\mathcal O(1)$)-region, the WDW behavior appears again away from the singularity, as soon as such a wave packet is considered. This picture is plotted in Fig. 4.
\begin{figure}
\includegraphics[height=2in]{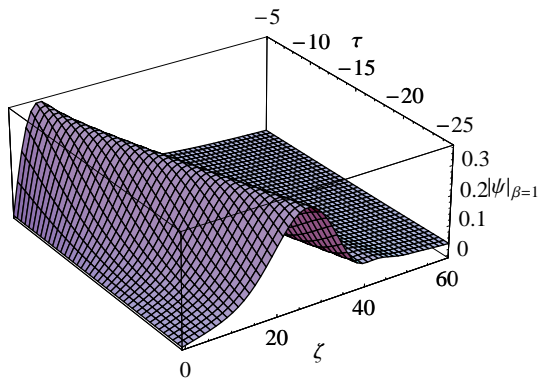} 
\caption{The evolution of the wave packets $\vert\Psi(\tau,\zeta)\vert$ for $\beta=1$ away from the singularity. The parameters are chosen as $k_0=0.001$ and $\sigma=1/4$.} 
\end{figure}

This behavior of the wave packets states the self-consistency of the GUP Taub Universe, since it reproduces the correct semiclassical dynamics, i.e. the WDW one previously described for localized states. In fact, it is only necessary the wave packets peak at energy much less than $1/\sqrt\beta$ to obtain again the WDW behavior far enough from the singularity. This way the consequences of GUP, i.e. the localization of the wave function near the potential wall at $x=x_0$, appear only at scales for which the deformation parameter $\beta$ can be appreciated, i.e. for $\beta p^2\sim \mathcal O(1)$. However, although the realization that the GUP effects on the quantum dynamics are important for $\beta p^2\sim \mathcal O(1)$ is clear since the deformed algebra (\ref{modal}), the fact that such effects become negligible far from the singularity is a non-trivial feature of the model.

\section{Concluding remarks}

In this paper we have shown the effects of a modified Heisenberg algebra, which reproduces a GUP as coming out from studies on string theory \cite{String}, on the Taub cosmological model. The dynamics of the model is analyzed in the ADM formalism. This procedure leads us to regard the variable $\tau$ as a time-coordinate for the system. Therefore only the physical degree of freedom of the Universe, which is connected to the Universe anisotropy, is quantized according to the GUP prescription while the time variable is treated canonically. 

The study was performed at both classical and quantum level. The classical analysis was carried out keeping the deformation parameter $\beta$ fixed as $\hbar\rightarrow0$, i.e. the parameter $\beta$ was regarded as independent with respect $\hbar$. What we find was that the deformed dynamical trajectories, before and after the bounce at the potential wall, are closer to each other with respect to the canonical case. At quantum level, the eigenfunctions were analyzed and the motion of the wave packets considered. The GUP theory was compared with the WDW approach, regarded as a limiting case of our analysis for $\beta=0$. In the WDW approach the wave packets follow the classical trajectory, after the bounce, toward the classical singularity. On the other hand, in the GUP approach, the differences on the motion of the wave packets for increasing values of $\beta$ are given in detail. In the limiting case when $\beta\sim\mathcal O(1)$ the Universe, from a probabilistic point of view, stands close to the potential wall and a stationary behavior is predicted. Therefore two main conclusions can be inferred:
\begin{itemize}
	\item The classical singularity is not probabilistically favored. As a matter of fact, the wave packets stop following the classical trajectories toward the singularity and a dominant peak (near the wall region) in the probability amplitude to find the Universe appears. In other words, the GUP Taub Universe exhibits a singularity-free behavior.
\end{itemize}

\begin{itemize}
	\item The large anisotropy states, i.e. those for $|\gamma_+|\gg1$, are probabilistically suppressed. In fact, the Universe wave function appears to be peaked at values of anisotropy $|\gamma_+|\simeq\mathcal O(10^{-1})$. In this respect, the GUP wave packets predict the establishment of a quantum isotropic Universe differently from what happens in the WDW theory. This feature is in agreement with the study performed in \cite{Vakili} on a more simple case (the $\gamma_-=0$ case of Bianchi I model), although these two approaches are substantially different.
\end{itemize}

{\bf Acknowledgments:} We would like to thank Alessandra Corsi for her help in re-styling the English of this work.

\end{document}